\documentclass[12pt]{article}
\usepackage{authblk}
\usepackage{caption}
\usepackage{subcaption}
\usepackage{booktabs}
\usepackage{amsmath}
\usepackage{graphicx}
\usepackage{tikz}
\usepackage{subcaption}
\usepackage{float}
\usetikzlibrary{shapes.geometric, arrows}
\usepackage{hyperref}
\usepackage{geometry}
\usepackage{amssymb}
\usepackage{epsfig}
\usepackage{amsfonts}
\usepackage{graphicx}
\usepackage{color}
\newcommand{\bit}{\begin{itemize}}
\newcommand{\eit}{\end{itemize}}

\bibliography{bib/references}

\begin{document}
\author[1]{Ion Bica}
\author[1]{Ryan Trang}
\author[1]{Rui Hu}
\author[1]{Wanhua Su}
\author[1]{Zhichun Zhai}
\author[2]{Qingrun Zhang}
\affil[1]{\centering {Department of Mathematics and Statistics, MacEwan University, 10700-104 Ave NW, Edmonton, AB, T5J 4S2, Canada.}
\newline
e-mail: bicai@macewan.ca, trangr@macewan.ca, hur3@macewan.ca, suw3@macewan.ca, zhaiz2@macewan.ca}
\affil[2]{\centering{Department of Biochemistry $\&$ Molecular Biology, University of Calgary, Calgary, Alberta, T2N 1N4, Canada.}

\centering{Department of Mathematics and Statistics, University of Calgary, Calgary, Alberta, T2N 1N4, Canada.}

\centering{Alberta Children’s Hospital Research Institute, University of Calgary, Calgary, Alberta, T2N 1N4, Canada.}
\newline
email: qingrun.zhang@ucalgary.ca}

\title{Learning Image Derived PDE-Phenotypes from fMRI Data }
\date{}

\maketitle
\begin{abstract}
Partial Differential Equations (PDEs) model various physical phenomena, such as electromagnetic fields and fluid mechanics. Methods like Sparse Identification of Nonlinear Dynamics (SINDy) and PDE-Net 2.0 have been developed to identify and model PDEs based on data using sparse optimization and deep neural networks, respectively. While PDE models are less commonly applied to fMRI data, they hold the potential for uncovering hidden connections and essential components in brain activity. Using the ADHD200 dataset, we applied Canonical Independent Component Analysis (CanICA) and Uniform Manifold Approximation (UMAP) for dimensionality reduction of fMRI data. We then used Sparse Ridge Regression to identify PDEs from the reduced data, achieving high accuracy in classifying attention deficit hyperactivity disorder (ADHD). The study demonstrates a novel approach to extracting meaningful features from fMRI data for neurological disorder analysis to understand the role of oxygen transport (delivery $\&$ consumption) in the brain during neural activity relevant for studying intracranial pathologies.
\end{abstract}



\section{Introduction}
Partial differential equations can describe a wide variety of phenomena in real life. For example, Maxwell's equations describe the electric and magnetic fields arising from distributions of electric charges and currents and how those fields change in time. Another example, Schrodinger's equation, is the fundamental postulate of Quantum Mechanics. Lastly, another important example, the Navier-Stokes equation in fluid mechanics, describes the flow of incompressible fluids. 

Researchers usually construct partial differential equations models based on physical laws and then estimate the model parameters by applying the observed data. However, while monitoring and recording data is quite common in science, building partial differential equations models from data is a highly nontrivial pursuit. Thus, an important question is how to learn the underlying partial differential equation that governs a particular phenomenon only using the data. The study of uncovering the underlying symbolical models from observations started by Bongard and Lipson \cite{Bongard} and then followed by Schmidt and Lipson \cite{Schmidt} These two pioneer works aim to compare numerical derivatives of experimental data with the analytic derivatives of candidate models. Schaeffer \cite{Schaeffer_1} developed a learning algorithm using sparse optimization to identify the terms in the underlying partial differential equations and approximate the terms' coefficients. This approach is known as the sparse identification of nonlinear dynamics (SINDy), which has been developed and applied in the literature recently, for example: Chang and Zhang \cite{Chang} Schaeffer et al. \cite{Schaeffer_2} Wu and Zhang \cite{Wu}

There are data-driven methods to discover the governing partial differential equations. To name some, Long, Lu and Dong \cite{Long} proposed a deep neural network method, PDE- Net 2.0. Unlike SINDy, PDE-Net 2.0 doesn't have to construct a dictionary of simple functions and partial derivatives that are likely to appear in the model, which usually leads to high memory load and computation cost.

The statistical and probabilistic approaches exploring functional magnetic resonance imaging (fMRI) data are widespread in scholarly works. However, the fMRI data are much less analyzed using the partial differential equations models approach, which are dominant tools for inferring hidden connections and searching essential components in many scientific disciplines. One of the reasons why fMRI data are much less analyzed using the partial differential equations models approach may be because of the need for more knowledge of the governing laws.

\section{fMRI and BOLD Signal}
The human brain has two types of matter: the gray matter, made up of cells bodies, which process sensations, control voluntary movements, and enable speech, learning and cognition, and the white matter, made up of axons (nerve fibres), which connect cells between them and project to the rest of the body. Thus, the white matter helps connect different brain regions and constitutes about half of the total human brain volume. 

Historically, scientists have focused on the gray matter in the cortex, thinking that that is the region where the action happens, ignoring the white matter.

Progressively, in recent years, researchers have started using functional magnetic resonance imaging (fMRI) to detect level-dependent signals of blood oxygenation (BOLD signal), a key marker of cerebral activity in the white matter, i.e., fMRI detects activation in white matter.

In Gonzalez-Castillo et al. \cite{Gonzalez-Castillo} researchers addressed the limitation of detailed reading from fMRI scans, which traditionally emphasized only a localized view of the brain functions, showing sparse activated regions in the brain in response to stimuli or to performing a task. Their findings highlighted the subtle details in fMRI scans beyond their traditional analyses. They also emphasized the pervasiveness of false negative fMRI and how the sparseness of fMRI scans was not a result of localized brain functions but rather a consequence of high noise and overly strict predictive response models. They challenged the traditional fMRI analyses, showing that under optimal noise conditions, fMRI activations extended well beyond areas of primary relationship to a task, and blood-oxygen-level-dependent signal changes correlated with task-timing appeared in over 95$\%$ of the brain for a simple visual stimulation plus attention control task. Moreover, they showed that response shape varied substantially across regions and that whole-brain parcellation based on those differences produced distributed clusters that were anatomically and functionally meaningful, symmetrical across brain hemispheres, and reproducible across subjects.

In Gawryluk et al. \cite{Gawryluk} researchers showed emerging evidence in fMRI detection of the BOLD signal in the white matter and explained issues regarding the detection of white matter BOLD signals. Later on, Gore et al. \cite{Gore} provided multiple explanations for the issues raised by Gawryluk et al. \cite{Gawryluk}, where the BOLD signal had been overlooked in previous research or wholly dismissed as an artifact.

In Schilling et al. \cite{Schilling} researchers reported that BOLD signals increased significantly in the white matter and throughout the brain when people performed a task, like finger movement, and their fMRI brain scans were collected. 

While there still is not known the meaning behind the increase in the blood oxygenation signal in the white matter, researchers agree it is essential to study the BOLD signal in the white matter as it may provide a better understanding of the connectivity disruption in the brain in neurological disorders. 

The main discoveries of Schilling et al. \cite{Schilling} were the following:

\begin{itemize}
    \item BOLD signals open a new view of the interconnection of the brain functions; the brain functions communicate via BOLD signal transmissions.
    \item The map of the brain functions is an interconnected network which rethinks the accepted idea of brain functions manifesting in a sparse and localized mode that could be responsible for false negative fMRI readings; the subtle inter-regional differences in the BOLD signal response shapes contain sufﬁcient information to produce functional parcellations of the brain activity.
\end{itemize}

In gray matter, the tissue traditionally studied with fMRI, BOLD signals reflect an increase in blood flow (and oxygen) in response to increased activity of nerve cells. The researchers from Schilling et al. \cite{Schilling} hypothesized that axons, or the glial cells that maintain the protective covering of myelin around them, may also use more oxygen than when the brain is active. These signals could be related to what happens in the gray matter. Also, researchers discovered that the signal changed even if nothing biological happened in the white matter, and it changed differently in different white matter pathways and was present in all white matter pathways, which was a unique discovery.

However, new reports evidence reliable detection of BOLD signals in white matter, posing whether white matter displays transmission of BOLD signals through structural pathways in response to stimuli.

Studies like Schilling et al. \cite{Schilling} found that gray matter and white matter display time-locked activations to multiple stimuli, and both tissues showed statistically significant signal changes for all investigated task stimuli. In addition, different regions showed very different BOLD signal changes to the same task, and a region could display different BOLD responses to different stimuli. Researchers realized that the main challenges regarding the sparseness of activations in fMRI scans could be the following:

\begin{itemize}
    \item Elevated noise levels,
    \item Overly strict predictive response models.
\end{itemize}

In their approach, Vanderbilt University Medical Centre researchers increased the signal-to-noise ratio by putting the person whose brain was scanned to repeatedly repeat a visual, verbal or motor task to establish a trend and averaging the signal over several different fibre paths of white matter.

One of the reasons there are limited studies on the signals in the white matter is that they have a lower energy than signals from the gray matter and are, therefore, harder to distinguish from the background noise of the brain.

To find out, researchers will continue to study the changes in the white matter signals previously detected in schizophrenia, Alzheimer's disease and other brain disorders. Through studies on animals and tissue analyses, they also aim to determine the biological basis of these changes.

In conclusion:

\begin{itemize}
    \item When noise is sufﬁciently low and the response model versatile enough, the BOLD fMRI signal can reveal activity in most brain regions. 
    \item The whole brain continuously works and adapts to anticipate and switch on in response to the environment.
\end{itemize}

 In this paper, our primary objectives are to derive Partial Differential Equations (PDEs) from an fMRI dataset. Subsequently, we aim to apply the selected PDE features to the classification of neurological disorders, and in this article we focus on the ADHD-200 dataset.

The results of our study indicate that the PDE features extracted using our proposed method have high classification accuracy. Specifically, we achieved accuracy rates exceeding 70\%, which underscores the effectiveness of our approach in identifying relevant patterns and connections within the brain activity data.

Section 3 of this paper delves into the details of our methodology. It outlines the steps involved in preprocessing the datasets, the techniques used for PDE extraction, and the classification algorithms implemented. By providing a comprehensive description of our methodology, we aim to offer insights into the robustness and applicability of PDE-based feature extraction in the context of neurological disorder classification. 

\section{Dimensionality Reduction and PDE Learning}
In this section, we will learn a partial differential equation (PDE) from the ADHD-200 dataset, which is a comprehensive collection of neuroimaging and phenotypic data aimed at advancing research in understanding Attention Deficit Hyperactivity Disorder (ADHD). This dataset is part of the ADHD-200 Global Competition. To learn an effective PDE, we will need to preprocess the dataset, including smoothing the data and reducing its dimensionality. Following preprocessing, we will employ sparse ridge regression to learn the PDE.

\subsection{Dataset Description}
The ADHD200 dataset \cite{Bellec} includes preprocessed fMRI data for 973 individuals, aggregated from multiple international imaging sites. The participants range in age from 7 to 27 years and include both individuals diagnosed with ADHD and typically developing controls. More specifically, the fMRI data is four-dimensional in the spatial dimensions of the scan and the temporal dimension. Phenotypic data such as age, sex, handedness, and ADHD diagnosis are also provided. We selected Nilearn's ADHD dataset, which is a forty-individual subset of the original ADHD200 dataset. This dataset was chosen due to being pre-formatted to work with the Nilearn library, with additional quality assurance being done; additionally the dataset is balanced, having twenty individuals in both the control and ADHD afflicted classes. The dataset is available directly through Nilearn's Python library. The phenotype of interest is whether the subject has ADHD, which is a binary label.

\subsection{Dimensionality Reduction and PDE Derivation}
Our first objective is to derive a partial differential equation from the fMRI dataset to uncover hidden connections and essential components in brain activity. To achieve this, we will first reduce the dimensionality of the dataset, smooth it, and transform the reduced data into a mesh. We will then apply the PDE-Find algorithm to learn the PDE from this transformed data.

\subsubsection{Dimensionality Reduction}
Broadly, the dimensionality reduction is achieved via the following steps (also shown in Figure \ref{Dimesionality Reduction}). More details about the techniques used are discussed in the Appendix.
\begin{itemize}
    \item Step 1. Canonical Independent Component Analysis (CanICA) \cite{Varoquaux} is used to identify twenty Regions of Interest (ROIs) of the brain from the ADHD200 dataset. Figure \ref{fig:roi_all} shows a plot of all twenty ROIs selected via CanICA. Twenty aggregated BOLD time series are then extracted by averaging over the regions of interest. Traditional implementations of Independent Component Analysis (such as FastICA) are not robust to mild data variation. Due to high inter-subject variability in fMRI data, this may result in misalignment of identified regions of interest. As such, Canonical Independent Analysis (CanICA) was used to identify ROIs aligned across all subjects.

\begin{figure}[!ht]
\centering
\includegraphics[width=6in]{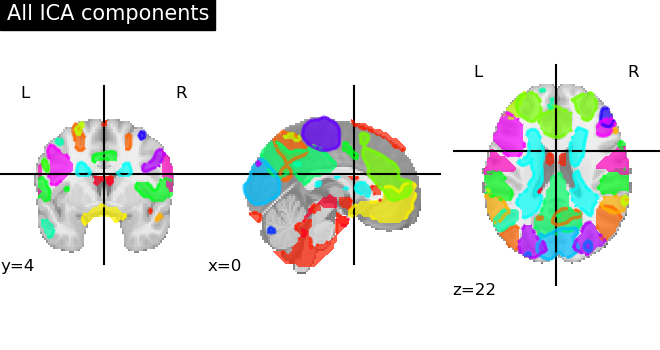}
\caption{All 20 identified ROIs via CanICA.}
\label{fig:roi_all}
\end{figure}
    
    \item Step 2.  Real Fast Fourier Transformation (RFFT) is applied to smooth BOLD data. RRFT is a Python function in the fundamental package for scientific computing in the Python library, NumPy, and it computes the one-dimensional discrete Fourier Transform for real input. The function computes the one-dimensional $n$-point discrete Fourier Transform (DFT) of a real-valued array by means of the Fast Fourier Transform (FFT). More about RFFT is explained in the Appendix.
    
    The smoothed BOLD data is shown in Figure \ref{fig:bold_rfft}. The Fourier-Filter (FFT Filter) is applied to smooth the data, i.e., the process of mapping a time-signal from time-space to frequency-space. This process gives us a better picture for how much of which frequency is in the original time-signal and we can ultimately filter-out some of these frequencies to remap back into time-space. Hence the surface generated by the model trained on the smoothed data is obtained.

\begin{figure}[!ht]
\centering
\includegraphics[width=4in]{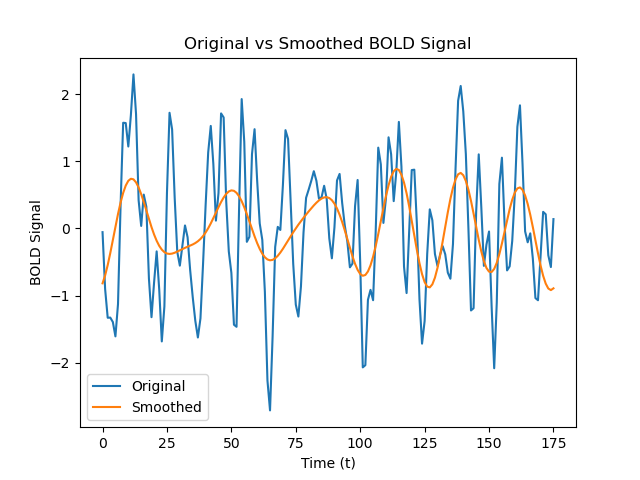}
\caption{Comparison of BOLD signal before and after RFFT smoothing.}
\label{fig:bold_rfft}
\end{figure}

    \item Step 3. Uniform Manifold Approximation and Projection (UMAP) \cite{McInnes} is a non-linear dimension reduction algorithm applied to the row-wise concatenated, smoothed BOLD data from all forty subjects. The technique learns the manifold structure of the BOLD data and finds a low-dimensional embedding that preserves the essential topological structure of that manifold (in our case, the spacetime structure). BOLD data generates high-dimensional data, and we use UMAP to provide a 2-dimensional representation, preserving the essential structures of the original dataset.
    \item Step 4. Once again, we apply the smoothing procedure to each of the two resulting time series obtained in Step 3.
\end{itemize}

\begin{figure}[!ht]
\centering
\begin{subfigure}[b]{0.48\linewidth}
\centering
\begin{tikzpicture}[
    node distance=1.5cm,
    startstop/.style={rectangle, rounded corners, minimum width=3cm, minimum height=1cm,text centered, draw=black, fill=red!30},
    process/.style={rectangle, minimum width=3cm, minimum height=1cm, text centered, draw=black, fill=orange!30},
    arrow/.style={thick,->,>=stealth}
]
\node (start) [startstop] {fMRI data};
\node (proc1) [process, below of=start] {CanICA};
\node (proc2) [process, below of=proc1] {RFFT Smooth};
\node (proc3) [process, below of=proc2] {UMAP};
\node (proc4) [process, below of=proc3] {RFFT Smooth};
\draw [arrow] (start) -- (proc1);
\draw [arrow] (proc1) -- (proc2);
\draw [arrow] (proc2) -- (proc3);
\draw [arrow] (proc3) -- (proc4);
\end{tikzpicture}
\caption{Dimensionality Reduction}
\label{Dimesionality Reduction}
\end{subfigure}
\begin{subfigure}[b]{0.48\linewidth}
\centering
\begin{tikzpicture}[
    node distance=1.5cm,
    startstop/.style={rectangle, rounded corners, minimum width=3cm, minimum height=1cm,text centered, draw=black, fill=blue!30},
    process/.style={rectangle, minimum width=3cm, minimum height=1cm, text centered, draw=black, fill=green!30},
    arrow/.style={thick,->,>=stealth}
]
\node (start) [startstop] {Dimensionality Reduction};
\node (proc5) [process, above of=start] {Train XGBoost Regressor};
\node (proc4) [process, above of=proc5] {Generate data $u(x, t) $};
\node (proc3) [process, above of=proc4] {Train Sparse Ridge Regression};
\draw [arrow] (start) -- (proc5);
\draw [arrow] (proc5) -- (proc4);
\draw [arrow] (proc4) -- (proc3);
\end{tikzpicture}
\caption{PDE Learning}
\label{PDE Learning}
\end{subfigure}
\caption{Workflow of the proposed method}
\end{figure}
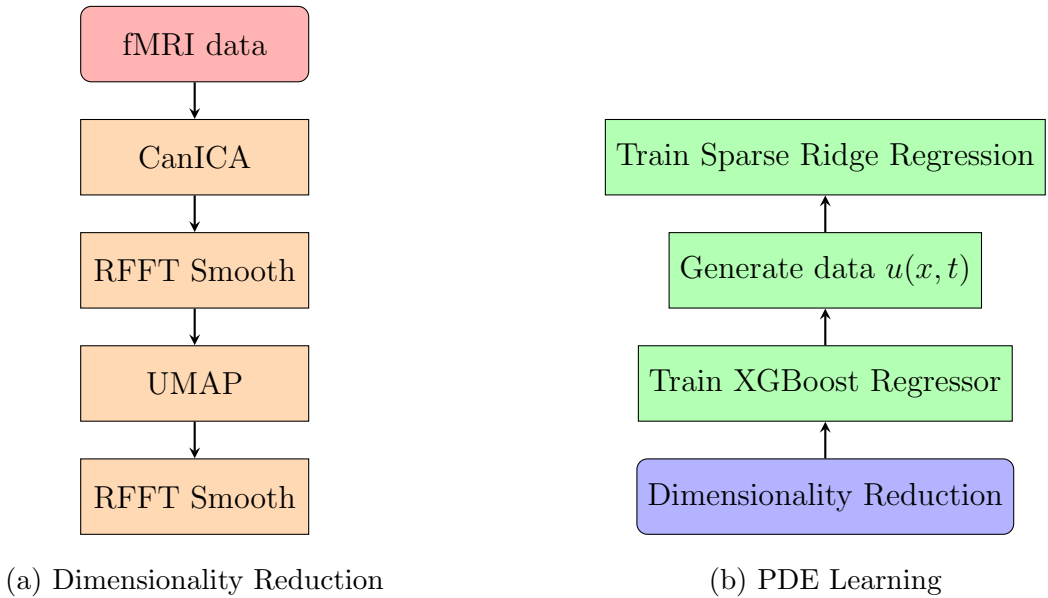

\subsubsection{Learning PDE}
In this section, we show steps to learn PDE from the preprocessed data as described in the previous section. The overview for learning the PDE coefficients from the embedding is illustrated in Figure \ref{PDE Learning}. 
\begin{itemize}
\item Step 1.  Via XGBoost Regressor \cite{Chen}  we learn the function $u(x, t) = y$ by training a regression model on $x$ and $t$ to predict $y$, and use the fitted model to generate all the points in our desired mesh. The mesh is generated this way since PDE-Find requires the values of $y$ from the mesh, and also the corresponding grid spacing of $x$ and $t$ values (i.e., $dx$ and $dt$) as inputs to learn the PDE.
    \item Step 2.  Sparse ridge regression is used to attempt to learn the PDE for each subject using their corresponding mesh. The library  that we use by Sparse Ridge Regression is $\{1, u, u_{xx}, u_{x}, uu_{xx}, uu_{x} \}$.
\end{itemize}

\subsubsection{Derived PDE Results}
We found that for 95 percent of samples (thirty-eight out of forty individuals, leaving two individuals atypical to the analysis due to the shape of their data sample) from the ADHD200 dataset, we were able to find a best-fit representation in terms of a PDE via Sparse Ridge Regression. This yielded a dataset of six PDE features (including the constant term) as illustrated in Table \ref{Sample of Learned PDE coefficients}, which shows a sample of five individuals from the ADHD200 data set, where Class 1 means they have ADHD, and Class 0
means they do not have ADHD.

A 95$\%$ bootstrap confidence interval for each PDE feature, using 10000 iterations (resamples), is provided in Table \ref{Bootstrap 95_1}. We found that two PDE features  (the coefficients of $u_{xx}$ and $uu_{xx}$) are statistically significant since their 95$\%$ bootstrap confidence intervals exclude zero. Table \ref{Sample of Learned PDE coefficients} suggests that individuals with ADHD tend to exhibit positive coefficients for $u_{xx}$ and negative coefficients for $uu_{xx}$ whereas individuals without ADHD show the opposite pattern, with negative coefficients for $u_{xx}$ and positive coefficients for $uu_{xx}$. 

\begin{table}[h]
\centering
\begin{tabular}{|c|c|c|c|c|c|c|c|c|}
\hline
$Subject ID$ & $constant$ & $u$ & $u_{xx}$ & $u_{x}$ & $uu_{xx}$ & $uu_{x}$ & $Class$   \\
\hline
1& 0.04677 & -0.001248 & 0.003257 & 0.055501 & -0.000860 & 0.000434 & 1 \\
2& 0.03228 & -0.000609 & 0.001092 & 0.014686 & -0.000149 & -0.003414 & 1 \\
3& 0.01998 & 0.000000 & 0.001277 & 0.000000 & -0.000274 & 0.000000 & 0  \\
4& -0.019349 & 0.005080 & -0.001478 & -0.026406 & 0.000476 & 0.007237 & 0 \\
5& -0.008139 & 0.000608 & -0.000811 & 0.000000 & 0.000202 & 0.001380 & 0 \\
\hline
\end{tabular}
\caption{Sample of Learned PDE coefficients.}
\label{Sample of Learned PDE coefficients}
\end{table}

\begin{table}[htbp]
\centering
\begin{tabular}{lccc}
\toprule
\multicolumn{4}{c}{PDE Features} \\
\midrule
Feature & 2.5\% & 97.5\% & Mean \\
\midrule
$constant$ & -0.254156 & 0.056553 & -0.086180 \\
$u$ & -0.005866 & 0.057578 & 0.023803 \\
$u_{xx}$ & 0.000416 & 0.025778 & 0.012976 \\
$u_{x}$ & -0.015969 & 0.084424 & 0.032278 \\
$uu_{xx}$ & -0.006522 & -0.000116 & -0.003235 \\
$uu_{x}$ & -0.13647 & 0.015140 & 0.001109 \\
\midrule
\end{tabular}
\caption{Bootstrap 95\% Confidence Intervals for PDE Features.}
\label{Bootstrap 95_1}
\end{table}

\subsection{Classification by Applying PDE Features}
\begin{figure}[!h]
\centering
\begin{minipage}{.5\textwidth}
  \centering
  \includegraphics[width=\linewidth]{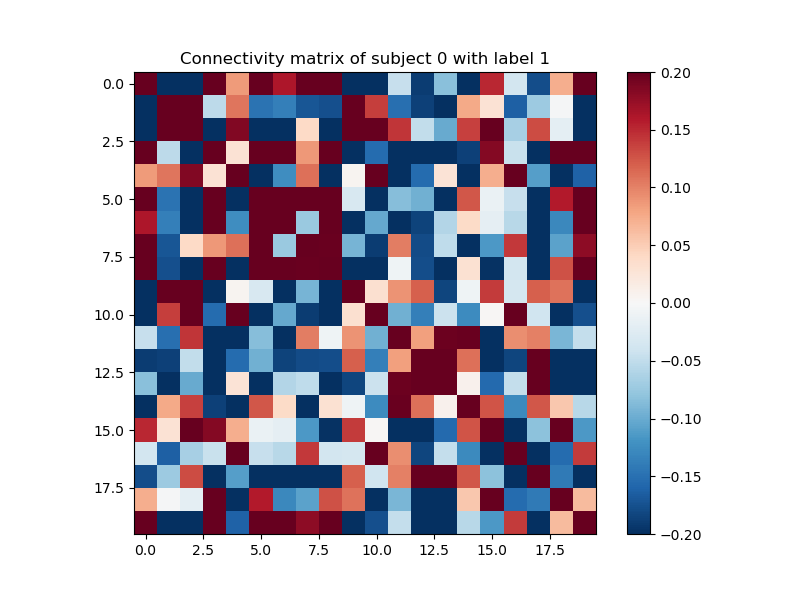}
  \caption{ROI Correlation Matrix}
  \label{fig:ROI_corr_matrix}
\end{minipage}%
\centering
\begin{minipage}{.5\textwidth}
  \centering
  \includegraphics[width=\linewidth]{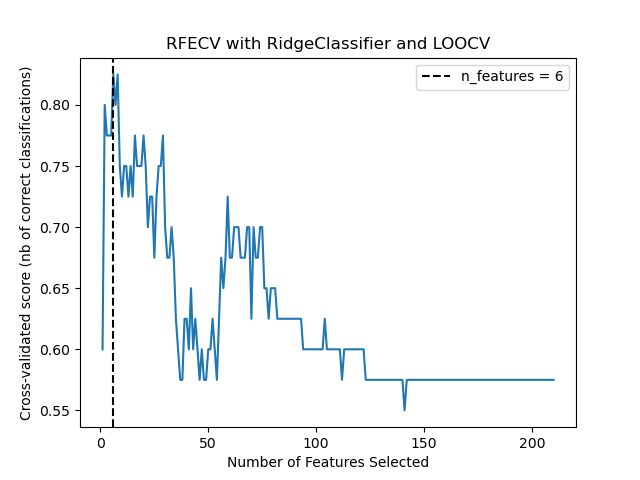}
  \caption{RFECV for Correlation Features. Six features were selected.}
  \label{fig:RFECV_corr_feats}
\end{minipage}%
\end{figure}

In this section, we apply the two significant PDE features for classification. The performance of these PDE features will be compared against the ROI correlation matrix features (Figure \ref{fig:ROI_corr_matrix}), which serve as a reference point for evaluating the classification performance of the PDE coefficients.

We used the same ROIs identified by CanICA for the PDE features but focused on the lower-triangular region of the ROI correlation matrix. For a sample with twenty components, this results in 210 features. Six ROI correlation matrix features, as shown in Table \ref{Bootstrap 95_2}, were selected using Recursive Feature Elimination \cite{Guyon}, as they are significant at the 95$\%$ confidence level obtained via bootstrapping (see Figure \ref{fig:RFECV_corr_feats}).

\begin{table}[htbp]
\centering
\begin{tabular}{lccc}
\toprule
\multicolumn{4}{c}{Correlation Matrix Features} \\

\midrule
Feature & 2.5\% & 97.5\% & Mean \\
\midrule
ROI correlation 103 & 0.178983 & 1.645006 & 0.941947 \\
ROI correlation 146 & 0.584835 & 1.863020 & 1.267607 \\
ROI correlation 154 & -1.426459 & -0.325181 & -0.897626 \\
ROI correlation 194 & 0.195966 & 1.265697 & 0.751748 \\
ROI correlation 197 & 0.700806 & 1.838815 & 1.292869 \\
ROI correlation 199 & -1.345755 & -0.239360 & -0.795475 \\
\bottomrule
\end{tabular}
\caption{Bootstrap 95\% Confidence Intervals for Correlation Matrix Features.}
\label{Bootstrap 95_2}
\end{table}

A Support Vector Machine (SVM) with a Radial Basis Function (RBF) kernel is used for classification. Two cross-validation schemes are used: Stratified K-Fold (with \(K=5\) for an 80-20 train-test split per fold) and Leave-One-Out Cross Validation (LOOCV). The results, shown in Table \ref{Cross Validation Results}, present the cross-validation outcomes using both the correlation matrix and PDE feature sets. The Support Vector Machine (SVM) model using correlation matrix features achieved a mean 5-fold cross-validated accuracy of 0.90, compared to 0.75 for the model with PDE features. Similarly, the mean leave-one-out cross-validated accuracy was 0.95 for the correlation matrix features, outperforming the 0.725 accuracy observed for the PDE features. In both cross-validation methods, the SVM model with correlation matrix features demonstrated superior performance over the model with PDE features.  

\begin{table}[!htbp]
\centering
\begin{tabular}{lcccc}
\toprule
& \multicolumn{2}{c}{Stratified K-Fold (K=5)} & \multicolumn{2}{c}{LOOCV} \\
\cmidrule(lr){2-3} \cmidrule(lr){4-5}
Feature Set & Mean CV Accuracy & Mean CV Log-Loss & Mean CV Accuracy & \\
\midrule
Correlation Matrix Features & 0.900 & 0.349 & 0.950 & \\
PDE Features & 0.750 & 0.590 & 0.725 & \\
\bottomrule
\end{tabular}
\caption{Combined Cross-Validation Results}
\label{Cross Validation Results}
\end{table}

\begin{figure}[!h]
\centering
\begin{minipage}{.5\textwidth}
  \centering
  \includegraphics[width=\linewidth]{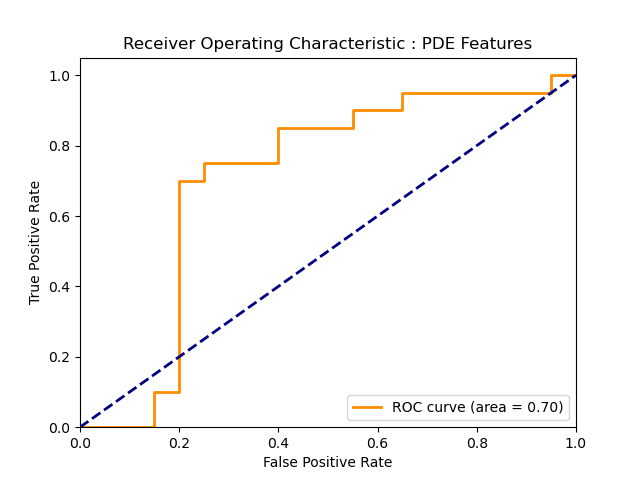}
  \caption{ROC of PDE Features}
  \label{fig:pdeconfmat}
\end{minipage}%
\centering
\begin{minipage}{.5\textwidth}
  \centering
  \includegraphics[width=\linewidth]{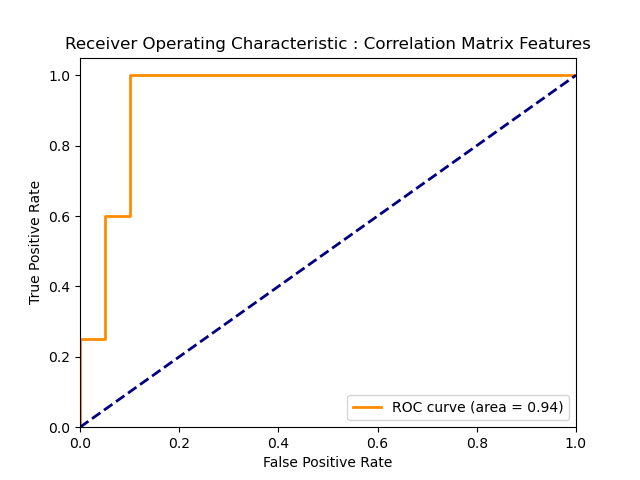}
  \caption{ROC of  Correlation Matrix Features}
  \label{fig:pdeconfmat2}
\end{minipage}%
\end{figure}

We also provide the Receiver Operating Characteristic (ROC) curves for the SVM classifiers fitted on the PDE features and correlation matrix features in Figure \ref{fig:pdeconfmat} and Figure \ref{fig:pdeconfmat2} respectively. The SVM with the correlation matrix features gives an area under the ROC curve (AUC) of 0.94, compared to 0.7 for the model with the PDE features. 

While Correlation Matrix features outperformed PDE features in classification accuracy as shown in the above results, our novel methodology enables the potential for further interpretability by characterizing subjects by their PDE coefficients. There is the potential for this data to be used as a new type of Image-Derived Phenotype (IDP) as mediator variables that provide systematic information of the functional relevance. 
Additionally, there is further potential for other researchers with expertise to extract insights about the PDEs.

\section{Proposed PDE Model}
BOLD-fMRI measures the resting-state functional connectivity in the brain at a whole-brain level, i.e., the temporal consistency of spontaneous neural activity between distinct regions in the brain. During the neural activity, the relative concentration of oxygenated and deoxygenated hemoglobin causes fluctuations in the T2/T2 FLAIR signal. At the same time, the BOLD response detects changes in the relative oxygen concentration in hemoglobin, thus making the BOLD response correlate to the T2/T2 FLAIR signal fluctuations. The T2/T2 FLAIR signal is essential in understanding intracranial pathologies \cite{Nicholson} thus making the study of properties and role of oxygen transport (delivery $\&$ consumption) in the brain during neural activity relevant for studying intracranial pathologies.

As Vazquez et al. explain \cite{Vazquez} transporting oxygen to the brain is a complex process. It is delivered to tissue at the capillary level by diffusion, a fundamental process in our research. In tissue, oxygen diffuses until cellular mitochondria are using it all up. This oxygen diffusion manifests by the oxygen concentration gradient induced between spatially distinct brain regions, an essential aspect of our study.

The fact that the significant features discovered in the PDE learning are both $u_{xx}$ and $uu_{xx}$ suggests that an appropriate PDE reaction-diffusion model describing the oxygen transport to the brain is:
\begin{equation}\label{PDE_Model}
   u_t=(D(u)u_x)_x
\end{equation}
where $u:=u(x,t)$ is the oxygen concentration in the brain [$\%O_2$], at spatial location $x$ [$cm$], at time $t$ [$s$], and the oxygen diffusivity function, $D(u)$, is a function depending on the oxygen concentration, $u$, as follows:
\begin{equation}
    D(u)=d_1+d_2u,~d_1,~d_2~{\rm constants}.
\end{equation}

In model (\ref{PDE_Model}), we considered the net rate at which oxygen is delivered and consumed to be in balance, i.e., the amount of cerebral oxygen delivered is consumed in equal amounts by the brain tissue, i.e., rate of oxygen delivery minus the rate of oxygen consumption is zero.

The constant $|d_1|$ represents the oxygenation diffusivity, and its units are [$cm^2/s$]. The constant $|d_2|$ represents the oxygenation diffusivity per oxygen concentration, and its units are [$cm^2/(s\cdot \%O_2)$].

Our proposed model (\ref{PDE_Model}) comprehensively describes the behaviour of oxygen concentration in the brain. It considers the competition between the diffusion of oxygen and the delivery of oxygen through cerebral blood flow. The oxygen diffusion manifests along the oxygen concentration gradient [$\%O_2/cm$], a crucial aspect of our model, which will help in creating the boundary conditions for the model (\ref{PDE_Model}). The initial condition imposed on the model (\ref{PDE_Model}) will be the percentage of oxygen carried through the cerebral arteries to the brain, providing a clear starting point for our research.

Using Table \ref{Sample of Learned PDE coefficients}, we show a scenario for brain oxygenation fluctuations for subjects $1,~2$ and $4,~5$. The model (\ref{PDE_Model}) was consider over a unit of space dimension $[0\,cm,~1\,cm]$ over the time interval $[0\,s,~1\,s]$, with assumed initial and boundary conditions as follows:
\begin{itemize}
    \item Initial Condition:
    \begin{equation}
        u(x,0)=0.95
    \end{equation}
    The initial condition signifies the initial resting saturation oxygen level, which was considered $95\%$.
    \item Boundary Conditions:
    \begin{equation}
        u_x(0,t)=-0.01,~u_x(1,t)=0.01
    \end{equation}
    The boundary conditions signify the oxygen gradient at $x=0$, and at $x=1$, respectively.
\end{itemize}

\begin{figure}[!ht]
\centering
\begin{minipage}{.25\textwidth}
  \centering
  \includegraphics[width=\linewidth]{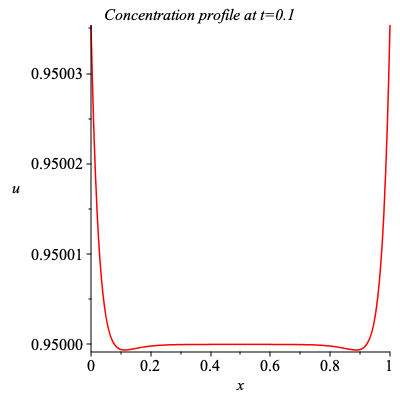}
\end{minipage}%
\begin{minipage}{.25\textwidth}
  \centering
  \includegraphics[width=\linewidth]{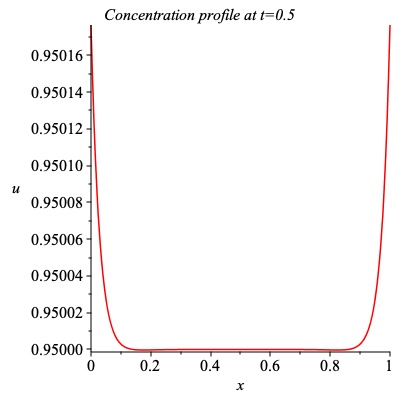}
\end{minipage}%
\begin{minipage}{.25\textwidth}
  \centering
  \includegraphics[width=\linewidth]{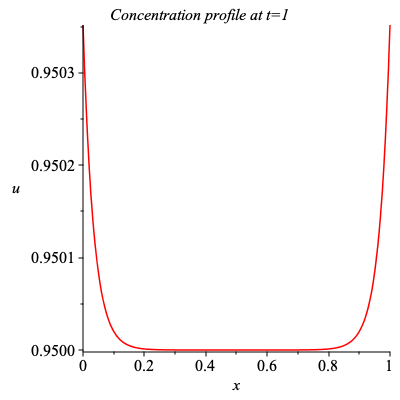}
\end{minipage}%
\centering
\begin{minipage}{.25\textwidth}
  \centering
  \includegraphics[width=\linewidth]{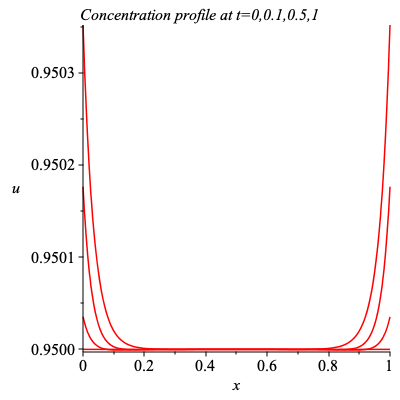}
\end{minipage}%
\caption{Oxygen concentration profiles for subject 1 in Table \ref{Sample of Learned PDE coefficients}}
\label{fig:subject_1}
\end{figure}

\begin{figure}[!ht]
\centering
\begin{minipage}{.25\textwidth}
  \centering
  \includegraphics[width=\linewidth]{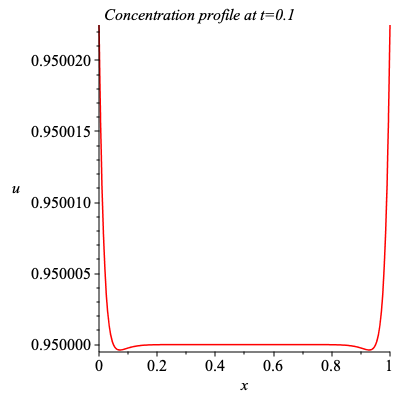}
\end{minipage}%
\begin{minipage}{.25\textwidth}
  \centering
  \includegraphics[width=\linewidth]{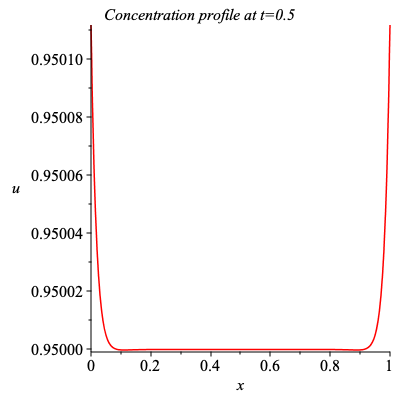}
\end{minipage}%
\begin{minipage}{.25\textwidth}
  \centering
  \includegraphics[width=\linewidth]{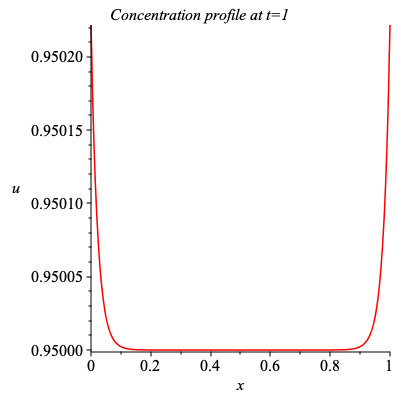}
\end{minipage}%
\centering
\begin{minipage}{.25\textwidth}
  \centering
  \includegraphics[width=\linewidth]{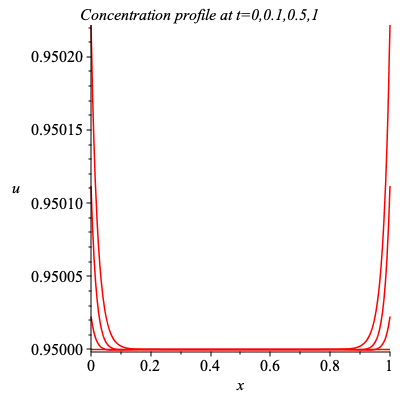}
\end{minipage}%
\caption{Oxygen concentration profiles for subject 2 in Table \ref{Sample of Learned PDE coefficients}}
\label{fig:subject_2}
\end{figure}

\begin{figure}[!ht]
\centering
\begin{minipage}{.25\textwidth}
  \centering
  \includegraphics[width=\linewidth]{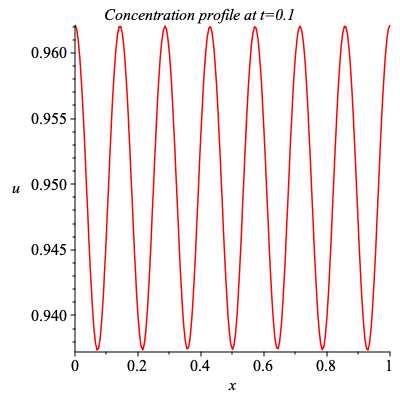}
\end{minipage}%
\begin{minipage}{.25\textwidth}
  \centering
  \includegraphics[width=\linewidth]{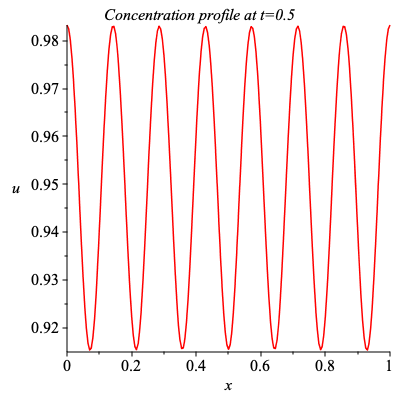}
\end{minipage}%
\begin{minipage}{.25\textwidth}
  \centering
  \includegraphics[width=\linewidth]{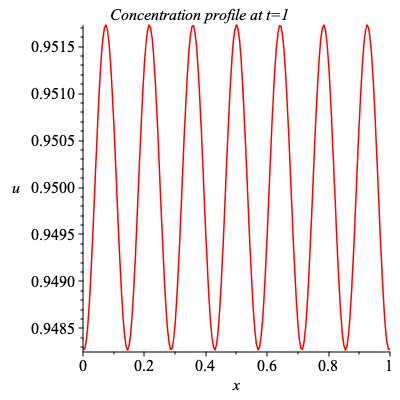}
\end{minipage}%
\centering
\begin{minipage}{.25\textwidth}
  \centering
  \includegraphics[width=\linewidth]{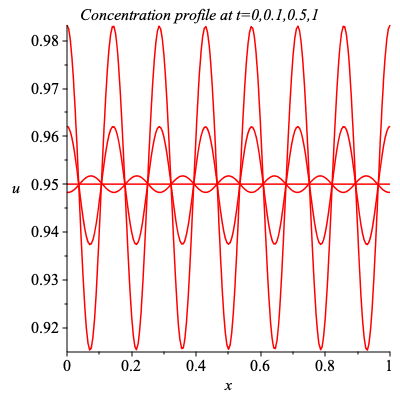}
\end{minipage}%
\caption{Oxygen concentration profiles for subject 4 in Table \ref{Sample of Learned PDE coefficients}}
\label{fig:subject_4}
\end{figure}

\begin{figure}[!ht]
\centering
\begin{minipage}{.25\textwidth}
  \centering
  \includegraphics[width=\linewidth]{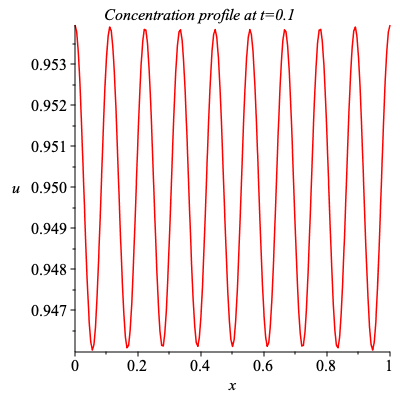}
\end{minipage}%
\begin{minipage}{.25\textwidth}
  \centering
  \includegraphics[width=\linewidth]{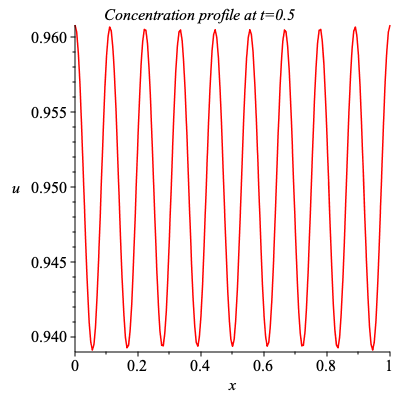}
\end{minipage}%
\begin{minipage}{.25\textwidth}
  \centering
  \includegraphics[width=\linewidth]{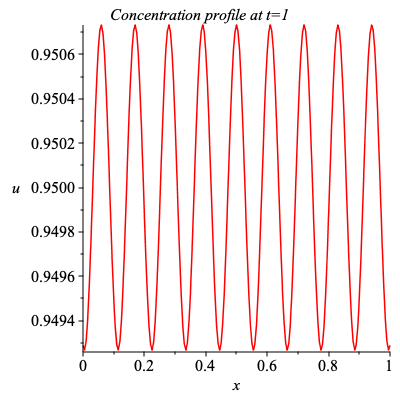}
\end{minipage}%
\centering
\begin{minipage}{.25\textwidth}
  \centering
  \includegraphics[width=\linewidth]{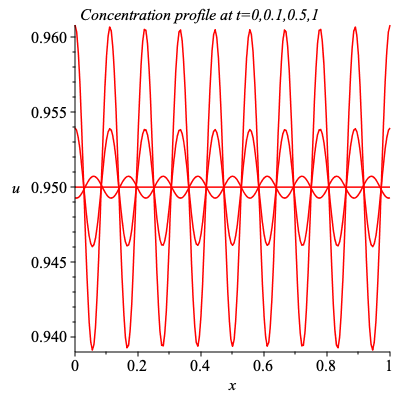}
\end{minipage}%
\caption{Oxygen concentration profiles for subject 5 in Table \ref{Sample of Learned PDE coefficients}}
\label{fig:subject_5}
\end{figure}

The subjects $1$ and $2$ have ADHD, and they display similar behaviour regarding the brain oxygenation fluctuations as illustrated in Figures \ref{fig:subject_1} and \ref{fig:subject_2}, respectively. As well, the subjects $4$ and $5$ do not have ADHD, and they display similar behaviour regarding the brain oxygenation fluctuations as illustrated in Figures \ref{fig:subject_4} and \ref{fig:subject_5}, respectively.
\newpage
\section{Summary}
This paper proposes a unified framework to extract important features from functional magnetic resonance imaging (fMRI) data that capture different brain functions or connectivity aspects. First, Canonical Independent Component Analysis is used to identify regions of interest (ROIs). The BOLD signal is smoothed for each ROI using a Real Fast Fourier Transform. Second, Uniform Manifold Approximation is applied to reduce the smoothed BOLD time series to two components. In the third step, XGBoost is used to learn the function $y=u(x, t)$ where $t$ is the time stamp. A surface is generated over a grid of $x$ and $t$ using the fitted XGBoost. Last, PDE-Find, which implements Sparse Ridge Regression, is used to learn a PDE from the generated surface $u(x, t)$. Two PDE components with coefficients that significantly differ from zero are regarded as important features. We applied the proposed PDE feature extractor to the pre-processed ADHD200 dataset and fed the important features to support vector machines. Both the classification accuracy and area under the ROC curve indicate that the identified PDE features are useful predictors for ADHD. 

The study opens new avenues for fMRI research, suggesting that the proposed PDE feature extractor could be applied to other neurological disorders like ADHD. The important features extracted by the proposed method not only achieve high classification accuracy but also provide insights into the underlying dynamics of brain activity, potentially improving diagnostic accuracy and therapeutic strategies. 

Future work could focus on expanding this approach to larger and more diverse datasets, as well as exploring the relationship among the extracted PDE features, the genotype, and the disease status. A deeper investigation into the interplay between PDE features and genetic information could provide more comprehensive insights into the underlying dynamic brain functionality and how these factors collectively influence disease progression and manifestation. 

Ultimately, this research represents a significant step towards integrating advanced mathematical modeling with practical medical imaging, promising enhanced insights and outcomes in neuroscience and clinical practice.

\section{Concluding Remarks and Future Work}
In this work, we proposed a novel method for extracting imaging-derived phenotypes (IDPs) from fMRI data based on the roles they played in a PDE. As far as our knowledge, state-of-the-art works all focused on particular regions or functions in the brain based on biological or medical knowledge \cite{Bongard,Chang,Long,Schaeffer_1,Schaeffer_2,Schmidt,Wu} In contrast, our method extracts IDPs de novo from PDE modelling. An obvious limitation might be the lack of immediate biological interpretations. However, this is also an advantage given the current limited understanding of human brains as well as the pathology of brain disorders. It is our expectation that the extracted IDPs may represent a higher level of “abstract” aggregation of system-level brain functions, revealing variations within brains that are functionally unknown. Indeed, extracting important however not previously known features will move the field of brain studies forward by breaking the boundary of our existing knowledge.  

To realize this promising impact, an immediate follow-up method development is to develop a statistical framework that leverages our PDE-informed IDPs to further biological discovery, particularly the causes of brain disorders. In connection to this, we are working on a new method that uses our new IDPs in mapping the genetic basis of brain disorders. In the field of genotype-phenotype association studies, there is an emerging trend that uses “mediators” in between of complex diseases and their genetic basis \cite{Brandes,Gamazon, Gusev} Towards this line, we have developed several novel statistical tools that leverage different types of mediators in between genotype and phenotype. These mediators include RNA gene expressions \cite{Cao,He_2,Kossinna} 3D genomes \cite{Li} transcription factors binding sites \cite{He_1} as well as brain images with IDPs extracted using standard methods \cite{He_3} However, all these methods are based on local features such as a gene expression or a region of the brain, without systematic information of the functional relevance. Now with the availability of the PDE-informed IDPs, we will be able to map genetic basis of brain disorders mediated by some complex (unknown) functions, unlocking the potential of using fMRI data to decipher genetic basis of brain disorders.

\section*{Appendix}
\subsection*{Canonical Independent Component Analysis (CanICA)}
We apply CanICA to select regions of interest. CanICA is used in various applications including neuroscience for analyzing brain imaging data and in other fields where multi-source data integration is essential. CanICA seeks to find components that are both independent within each dataset and maximally correlated across datasets. 

We utilized the Nilearn library and performed CanICA with standardization, de-trending, high-pass filtering, low-pass filtering, and spatial smoothing using an isotropic Gaussian filter kernel with full width at half maximum (FWHM) enabled (i.e., the width of the Gaussian filter, expressed as the diameter of the area on which the filter value is above half its maximal value).

\subsection*{Real Fast Fourier Transformation (RFFT)}
In our methodology, RFFT is applied for data smoothing. RFFT is a specialized variant of the Fast Fourier Transform (FFT) algorithm designed to efficiently compute the Fourier transform of real-valued input data. RFFT is particularly suitable for applications in signal processing, where data often consists of real numbers, such as audio signals, time-series data, and image processing. 

Given a real-valued discrete input signal \( x[n] \) of length \( N \), the Real Fast Fourier Transformation (RFFT) computes its spectrum \( X[k] \), where \( k \) represents frequency bins.

The RFFT of \( x[n] \), denoted as \( X[k] \), is defined as:

\[
X[k] = \sum_{n=0}^{N-1} x[n] \cdot e^{-i2\pi \frac{kn}{N}}
\]
where \( i \) is the imaginary unit, \( k \) is the frequency bin index, and \( N \) is the length of the signal.

Given the spectrum \( X[k] \) obtained from the RFFT, the IRFFT computes the real-valued signal \( x[n] \) as:

\[
x[n] = \frac{1}{N} \sum_{k=0}^{N-1} X[k] \cdot e^{i2\pi \frac{kn}{N}}
\]
where \( n \) is the time-domain sample index, \( k \) is the frequency bin index, and \( N \) is the length of the signal. 

Furthermore, let $C$ be the cutoff index where all frequency components greater than or equal to $C$ are set to zero. Therefore the smoothing procedure is:
\[
x[n]_{smooth} = \frac{1}{N} \sum_{k=0}^{N-1} X'[k]\cdot e^{i2\pi \frac{kn}{N}}
\]
with $X'[k]$ defined by:
\[
X'[k] = 
\begin{cases} 
    X[k] & \text{if } k < C \\
    0 & \text{if } k \geq C \\
\end{cases}
\]

Suppose $T_{i}$ is the length of the time series for the $i$th subject ($S_{i}$), and $F \in [0, 1]$ is the parameter controlling the number of non-zero frequency components, then $C = \lceil{T_{i} \times F}\rceil$. In our results, we use $F = 0.1136$. 

\subsection*{Uniform Manifold Approximation (UMAP)}
UMAP is a powerful dimensionality reduction technique designed to maintain the global and local structure of high-dimensional data when projecting it into lower dimensions. UMAP operates by constructing a high-dimensional graph representing the data, which is then optimized for a low-dimensional graph that preserves the topological relationships. 

We fit UMPA on the row-wise concatenated smoothed BOLD data of all forty subjects to reduce the dimension from twenty to two features (components), and then we transform each subject individually with the fitted model.

This method is highly efficient, scalable, and can be applied to various types of data, including numerical, categorical, and text data. UMAP is particularly useful for visualizing complex datasets, clustering, and as a preprocessing step for other machine learning algorithms.

\subsection*{XGBoost Regressor and Mesh Data Generation}
By employing the XGBoost Regressor \cite{Chen}, we model the function \( u(x, t) = y \) by training the regression model on the variables \( x \) and \( t \) to predict \( y \). The trained model is then used to generate all the points within the desired mesh.

For each subject, we use the values of $x$, and its indices $t$ as training data for an XGBoost Regressor \cite{Chen}, and fit the model on the values of $y$ to learn the function $u(x, t) = y$.

We then generate the data mesh used by PDE-Find to attempt to learn a PDE using the fitted XGBoost regressor. The grid spacing $dx$ and $dt$ of the mesh $x$ and $t$ values is simply:
\[
dx = \frac{max(x) - min(x)}{|T_{i}|}  \thickspace  \thickspace \thickspace \hbox{and}  \thickspace \thickspace \thickspace dt = 1,
\]
where $T_i$ is the length of the time series for subject $i$. For a given subject, we predict $u(x, t)$ on the desired mesh.

\subsection*{PDE Identification via Sparse Ridge Regression}
We use PDE-Find \cite{Rudy}, which implements Sparse Ridge Regression to attempt to learn a PDE from the generated mesh data $u(x,t)$ obtained in the previous section.  

Let $\Phi$ be the library of terms used by Sparse Ridge Regression to learn a PDE of up to $K$ terms. In the case of a second-order non-linear PDE with polynomial degree at most one, $\Phi$ has $K = 7$ terms
\[
\Phi = \{ 1, u, u_{xx}, u_{x}, uu_{xx}, uu_{x}\}.
\]
Let $\xi$ be the corresponding vector of PDE coefficients.
We learn the PDE by the objective function:
\[
\xi = \arg\min_{\xi} \{ \lVert \Phi\xi - \hat{u} \rVert_2^2 + \lambda \lVert \xi \rVert_2^2 \}
\]
where $\hat{u}$ is a vector containing all the mesh data and $\lambda$ is the regularity parameter chosen by users.
\end{document}